\documentclass[12pt,preprint]{aastex}
\newcommand{\alfven}{Alfv\'{e}n}
\newcommand{\alfvenic}{Alfv\'{e}nic}
\newcommand{\pref}{\protect\ref}

\begin{document}

\shorttitle{Spicules, \alfven{} Waves and TR Structure}
\shortauthors{McIntosh, De Pontieu \& Tarbell}
\title{Reappraising Transition Region Line Widths in light of Recent \alfven{} Wave Discoveries}
\author{Scott W. McIntosh\altaffilmark{1,2}, Bart De Pontieu\altaffilmark{3}, Theodore D. Tarbell\altaffilmark{3}}
\email{mcintosh@boulder.swri.edu, bdp@lmsal.com, tarbell@lmsal.com}
\altaffiltext{1}{Department of Space Studies, Southwest Research Institute,1050 Walnut St, Suite 300, Boulder, CO 80302}
\altaffiltext{2}{High Altitude Observatory, National Center for Atmospheric Research,P.O. Box 3000, Boulder, CO 80307}
\altaffiltext{3}{Lockheed Martin Solar and Astrophysics Lab, 3251 Hanover St., Org. ADBS, Bldg. 252, Palo Alto, CA  94304}

\begin{abstract}
We provide a new interpretation of ultraviolet transition region emission line widths observed by the SUMER instrument on the Solar and Heliospheric Observatory ({\em SOHO}). This investigation is prompted by observations of the chromosphere at unprecedented spatial and temporal resolution from the Solar Optical Telescope (SOT) on {\em Hinode} revealing that all chromospheric structures above the limb display significant transverse (\alfvenic{}) perturbations. We demonstrate that the magnitude, network sensitivity and apparent center-to-limb isotropy of the measured line widths (formed below 250,000K) can be explained by an observationally constrained forward-model in which the line width is caused by the line-of-sight superposition of longitudinal and \alfvenic{} motions on the small-scale (spicular) structures that dominate the chromosphere and low transition region.
\end{abstract}


\section{Introduction}
\citet{Alfven1947} speculated that the broadening of coronal emission lines was the signature of \alfven{}
waves of sufficient strength to be a potential heating source for Edl\'{e}n's recently observed hot solar corona \citep[][]{Edlen1943}. The physical origin of the enhanced UV (and EUV) transition region (TR) and coronal line widths, and the amount of ``hidden'' energy that they represent, has been an open topic of debate, conflicting interpretation and conjecture in the community since. Interest in \alfven{} waves in this regard is still strong \citep[e.g.,][]{Chae1998, Erdelyi1998, Banerjee1998, Moran2003, Peter2003, OShea2005} owing to the recent numerical models of quiet coronal heating and fast solar wind acceleration depend critically on the presence of a significant energy carried by them \citep[e.g.,][]{Suzuki2005, Cranmer2005, Verdini2007}. Only one thing has prevented the validation of the observational and theoretical inferences; the direct observation of \alfven{} waves in the solar atmosphere. Recent unequivocal observations of low-frequency ($<$5mHz) propagating \alfvenic{} motions in the solar corona \citep[][]{Tomczyk2007} and chromosphere \citep[][]{DePontieu2007b}, plus the relationship of the latter with finely structured spicules observed at the solar limb \citep[][]{DePontieu2007a} with the Solar Optical Telescope \citep[SOT;][]{SOT} on {\em Hinode} \citep[][]{Hinode}, have inspired us to look again at the relationship between non-thermal emission line widths and spatially unresolved \alfvenic{} motions\footnote{For a discussion on why we call these motions \alfvenic{} and not MHD kink-mode waves, see the supporting online material of \citet{DePontieu2007b}}. A reappraisal of TR line widths is clearly necessary. Much of the previous scientific effort in this area, prior to the launch of the Solar and Heliospheric Observatory \citep[{\em SOHO};][]{Fleck+others1995}, is expertly summarized in Ch.~5.2 of \cite{Mariska1992}, but we will focus our discussion on the measurements of the Solar Ultraviolet Measurement of Emitted Radiation \citep[SUMER;][]{Wilhelm1995} instrument.

We present a new analysis of SUMER full-disk spectroheliograms taken in early 1996 \citep[e.g.,][]{Peter1999a,Peter1999b} paying close attention to the spatial variation of the line intensities and non-thermal widths. We compare the results of the data analysis with those of an observationally constrained forward model based on the relationship of chromospheric flows and \alfven{} waves that occur on spatial structures that are considerably smaller than the SUMER spatial resolution (1\arcsec). The synthesis of data and model analysis demonstrates that the superposition of transverse and longitudinal motions on sub-resolution structure can explain the observed magnitude, network sensitivity and center-to-limb isotropy of the measured non-thermal line broadening in cooler ($<$250,000K) TR lines. The last finding leads us to the conclusion that previous arguments made against the presence of  \alfven{} waves in the TR, based on the apparent isotropy of the line width variation \citep[e.g.,][]{Chae1998}, are invalid.

\section{SUMER Observations \& Data Analysis}\label{secobs}
We analyze full disk spectroheliograms in emission lines of \ion{C}{4} (1548.20\AA{}, $\log_{10} T_{e}=5.0$) and \ion{Ne}{8} (770.42\AA{}, $\log_{10} T_{e}=5.8$) acquired by SUMER in February 1996. The slit dimensions, step sizes and exposure times of each of these raster scans are documented in Table~1 of \citet[][]{Peter1999a}. The data are reduced following the main SUMER data reduction procedure. Like \citet[][]{Peter1999a, Peter1999b} we use a 4$\times$4 pixel spatial binning of the spectra so that the resulting ``spectroheliograms'' have 4\arcsec{} pixels, similarly we employ a Genetic Algorithm \citep[e.g., ][]{McIntosh1998} to fit the spectral profiles with a Gaussian. Each pixel in the spectral image then has a line (peak) intensity, position and $1/e$ width $w_{1/e}$ (measured in spectral pixel units). The SUMER instrumental width ($w_{inst}$) is determined using the FWHM of the fitted Gaussian ($= 2\sqrt{ln(2)}\;w_{1/e}$) and the SUMER routine {\tt CON\_WIDTH\_FUNCT\_3}\footnote{We note that the instrumental width calculation was incorrectly performed in \citet{Peter1999a, Peter1999b} and, as a result, the published values of $v_{1/e}$ for \ion{He}{1}, \ion{C}{4} and \ion{Ne}{8} are systematically too small by a factor of $\sim2\sqrt{ln(2)}$. This has been verified by private communication with H. Peter.}. We convert $w_{1/e}$ and $w_{inst}$ to km/s via the multiplying factor $(D_{\lambda}/\lambda_{0})c$, where $\lambda_{0}$ is the rest wavelength of  the line, $D_{\lambda}$ is the the spectral pixel scale of SUMER detector A at that wavelength \citep[see, e.g., Table~2 of][]{Wilhelm1995} and $c$ is the speed of light. Finally, the non-thermal line width $v_{1/e}$ $(= w_{1/e}^2 - w_{inst}^2 - w_{th}^2$) is computed using the thermal width $w_{th} (= \sqrt{2k_{B}T_{e}^{\ast} / m_{ion}}$ for an ion of mass $m_{ion}$ and peak formation temperature of $T_{e}^{\ast}$) assuming that the ion and electron temperatures are equal.

\begin{figure}
\epsscale{1.0}
\plotone{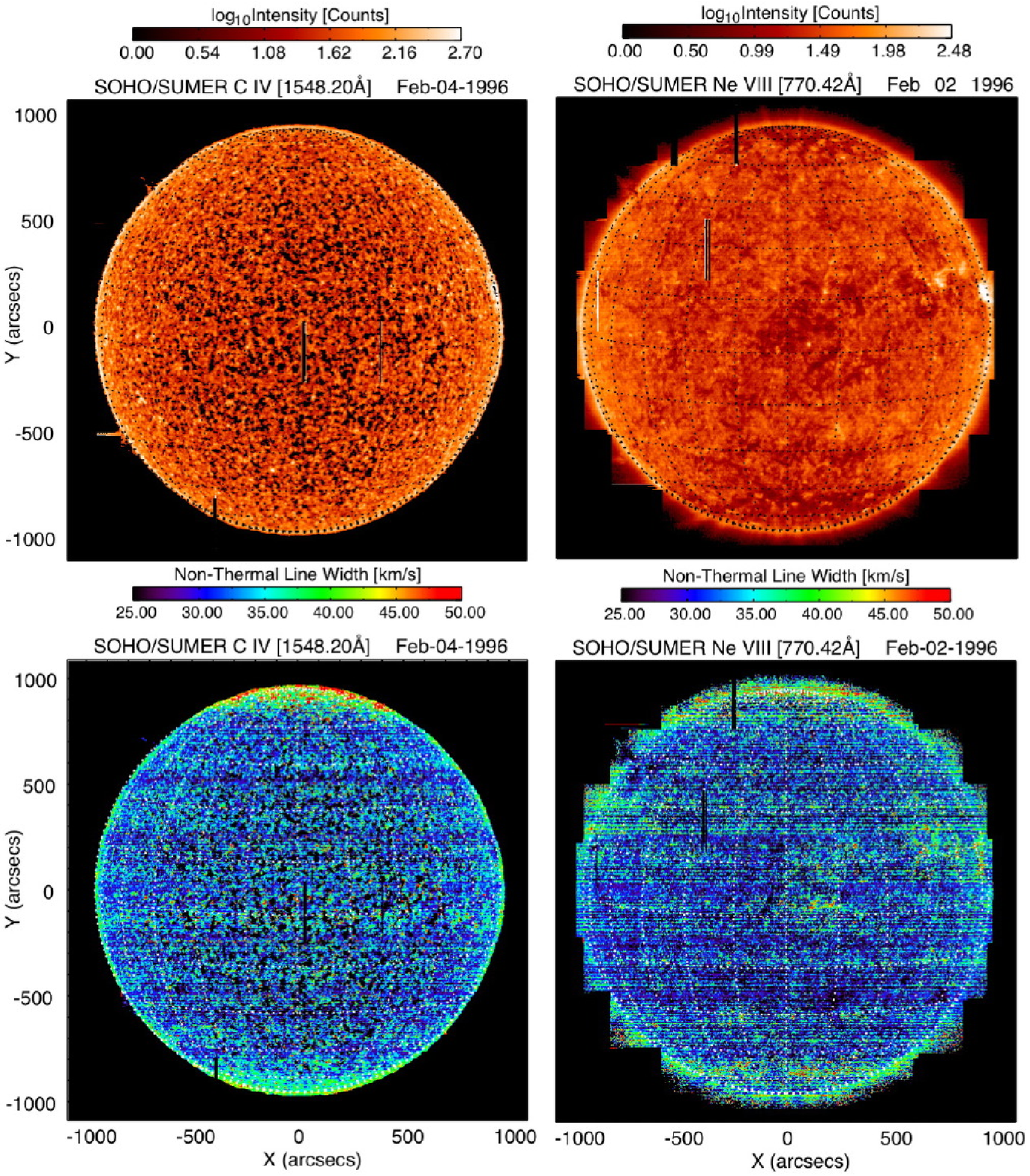}
\caption{SOHO/SUMER full disk spectroheliograms in emission lines of \ion{C}{4} (1548.20\AA{}; left) and \ion{Ne}{8} (770.42\AA{}; right). The top row shows the map of peak line intensity while the bottom row shows the map for the respective non-thermal line widths. Note the enhanced line widths in the bright supergranular network (left), around the limb (both) and from polar (both) and equatorial coronal hole (right) regions. Black vertical striping in the images masks corrupted spectral data. \label{f1}}
\end{figure}

Figure~\pref{f1} shows maps of the line peak intensities (top) and line width\footnote{From this point on we will simply refer to the non-thermal line width as the line width.} (bottom) from the GA fits for the SUMER spectra of \ion{C}{4} and \ion{Ne}{8}. The \ion{C}{4} map shows that the $v_{1/e}$ values are often enhanced by $\sim$10km/s around network (NW) locations compared to internetwork (IN) areas (note that some internetwork locations do not have sufficient counts to reliably fit a Gaussian to the spectra), an effect noted by \citet{Dere1984} among others. We also note that the non-thermal line broadening increases towards the limb for \ion{C}{4} and \ion{Ne}{8}. Also, the values of $v_{1/e}$ are enhanced (significantly) in polar coronal holes and (moderately) in the darker regions of the \ion{Ne}{8} spectroheliograms on the disk that are possibly the locations of equatorial coronal holes (ECH). Unfortunately, the poor spatial resolution and raster stepping of the spectroheliograms significantly limits the analysis of this interesting correspondence, leaving further investigation for more detailed SUMER measurements of ECH regions \citep[like that discussed by][]{McIntosh2007}.

\begin{figure}
\plotone{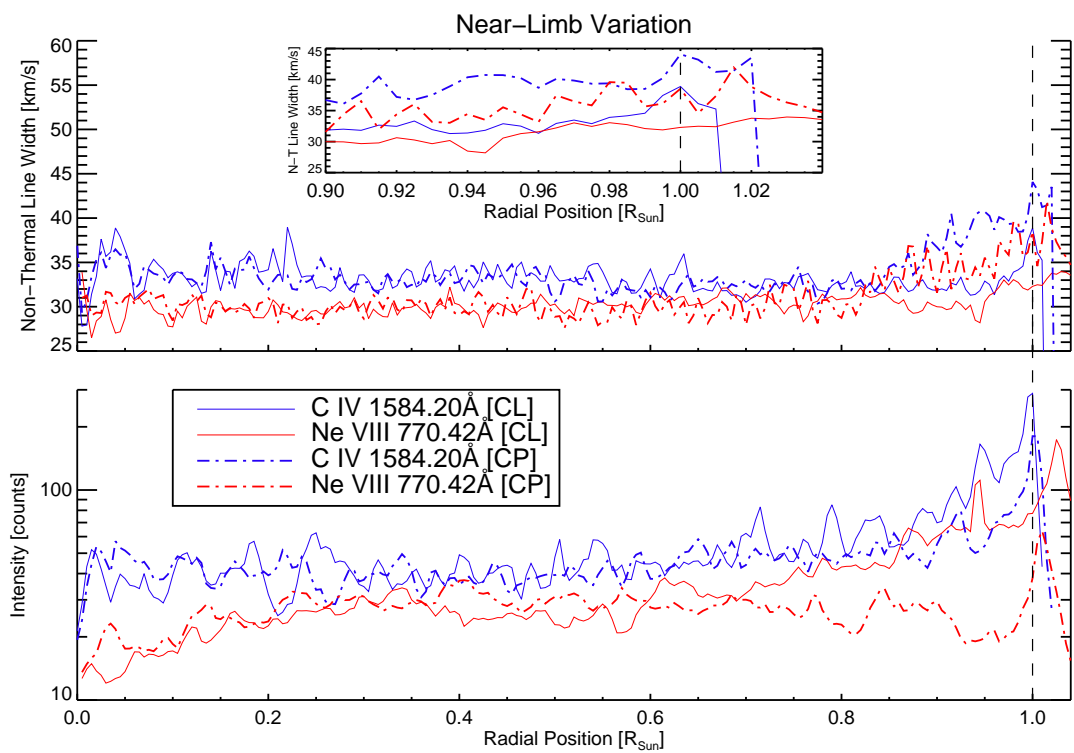}
\caption{Profiles of non-thermal line width (top) and peak intensity (bottom) in \ion{C}{4} (blue) and \ion{Ne}{8} (red) from SUMER spectroheliograms from the center-to-limb [CL; thin solid lines] and center-to-pole [CP; thick dot-dashed lines]. Each profile is computed using a 200\arcsec{} wide slice of the spectroheliogram for pixels with a chi-squared value less than two. The inset figure in the top panel highlights the variation of the line widths from 0.9 to 1.05 R$_{\sun}$. \label{f3}}
\end{figure}

Figure~\pref{f3} provides a detailed look at the center\--to\--limb (CL) and center\--to\--pole (CP) variation of the radially averaged line widths (top) and intensities (bottom) in both spectral lines for 200\arcsec{} wide swathes through disk center. In each case the line widths vary slowly from disk center to about a radial position of 0.9R$_{\sun}$ (or 0.8R$_{\sun}$ in the CP variation because of the polar coronal holes). Beyond 0.8-0.9R$_{\sun}$ a dramatic increase of 5-10km/s in the \ion{C}{4} line widths is visible, peaking around the limb and then dropping off to disk center values by about 10\arcsec{} above the limb \citep[noted by][]{Peter2003}. We note that the growth of the line widths in \ion{C}{4} is amplified in polar coronal holes \citep[also noted by][]{Peter1999a} and that the line width enhancements in the supergranular network (visible in Fig.~\ref{f1}) is reduced by the significant spatial averaging (cf. the narrow data slice shown in Fig.~\pref{f4}).

The \ion{Ne}{8} line width largely follows the pattern of \ion{C}{4} up to about 0.9R$_{\sun}$, but at the limb it shows a completely different variation. While the intensities peak somewhat above the limb ($\sim$1.03R$_{\sun}$) the line widths appear to increase very gradually beyond that point, extending some 40\arcsec{} above the limb as is evident from close inspection of the equatorial region in the appropriate panels in Fig.~\pref{f1}. The CP behavior of the \ion{Ne}{8} line width does not exactly follow that of the CL given its large growth inside the limb (similar to \ion{C}{4} CP) and gradual (uneven) increase above \citep[see, e.g.,][]{Banerjee1998}. The lack of growth above the polar limb is probably due to spatial averaging and the correspondence between polar plumes and significantly reduced line widths \citep[see, e.g.,][]{Wilhelm1998a} that is most clearly visible in the southern portion of the \ion{Ne}{8} Fig.~\pref{f1} panels. 

\section{Forward Model based on SOT Results}\label{data}
 
\begin{figure*}
\epsscale{1.0}
\plotone{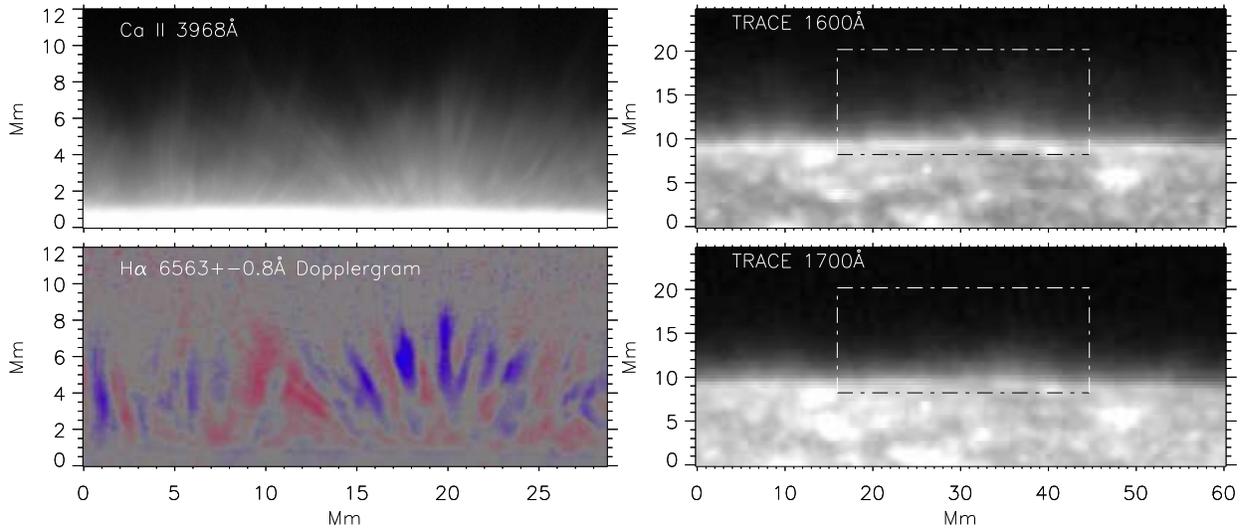}
\caption{Simultaneous {\em Hinode}/SOT \ion{Ca}{2}-H 3968\AA{} image, H$\alpha (\pm800$m\AA) dopplergram of the north pole coronal hole, and the lower-resolution TRACE 1600\AA{} and 1700\AA{} images of the same region (dot-dashed box in TRACE images is FOV of left panels). All images taken within 5 seconds of 15:05:30 UT on 27-Sep-2007. See the online edition of the Journal for a movie of this figure. \label{f2}}
\end{figure*}

We have seen that cool TR lines have a very different CL line width variation from those of the upper TR (or corona). We believe that this is because much of the lower TR is dominated by the extension of the chromospheric spicules observed with unprecedented spatial (0.2\arcsec) and temporal (5 s) resolution by {\em Hinode}/SOT. The left column of Fig.~\ref{f2} illustrates how the limb chromosphere is dominated by the superposition of a large number of long, thin ($\le$200km) spicules that extend to heights of between 5 and 10\arcsec{} and that undergo significant (at least $20-30$km/s) motions along and perpendicular to the line-of-sight (LOS). \citet{DePontieu2007a} demonstrates that there are two types of spicules present in the {\em Hinode} limb observations with very different dynamic properties. Type I spicules evolve on timescales of 3-5 minutes reaching velocities of 10-40km/s along their long axis \citep{DePontieu2007c}. Type II spicules on the other hand, occur on timescales of 10-60s and reach apparent longitudinal velocities of 50-150km/s. \citet{DePontieu2007b} have shown that these chromospheric features undergo vigorous \alfvenic{} motions with amplitudes around $20(\pm5)$km/s and periods ranging from 100 to 500s.


Why do we believe the same fine-scale structures show up in TR emission lines? A large body of literature exists describing the presence of TR spicules at the UV limb \citep[see, e.g.,][]{Mariska1992,Sterling2000,Wilhelm2000}. In addition, \citet{DePontieu2007a} find evidence for rapid heating of type II spicules to TR temperatures, a picture that is supported by our observations from the Transition Region and Coronal Explorer \citep[{\em TRACE};][]{Handy1999}. For example, Fig.~\pref{f2} (and the supplied online movie) shows that there is a qualitatively good correspondence between spicules visible in the {\em Hinode} \ion{Ca}{2}-H image and {\em TRACE} 1600\AA{} and 1700\AA{} passbands despite the gulf in spatial resolution. It is worth noting that off-limb the {\em TRACE} 1600\AA{} and 1700\AA{} passbands are dominated by \ion{C}{4} ($\sim10^{5}$K) and \ion{Fe}{2} ($\sim10^{4}$K) emission respectively, where the relative contributions are computed by folding the off-limb Skylab spectral atlas data \citep{Cohen1981} into the TRACE UV filter response functions (the 1600 passband contains 63\% \ion{C}{4} emission, and 27\% \ion{Fe}{2}, while the 1700 passband contains 62\% \ion{Fe}{2} and 18\% \ion{He}{2}). These findings strongly suggest that the lower TR is structured by the same spicules that \citet{DePontieu2007a} observe with SOT in the chromosphere \-- results compatible with inferred fine-scale structuring in the TR \citep[e.g.,][]{Dere1987}. Therefore, we deduce that lower TR dynamics are dominated by the same transverse and longitudinal spicular motions. Of course, the superposition of many finely-structured features undergoing vigorous motion naturally leads to non-thermal line broadening of emission lines observed by a UV spectrograph with coarse spatial resolution, like SUMER. \citet{Mariska1992, Doyle2005, Giannikakis2006} have previously proposed that flows {\em along} dynamic spicules are responsible for the increase of line widths at the limb and the former has attempted to model this phenomenon with varying degrees of success \citep[e.g.,][]{Mariska1976}. Here we provide, for the first time, a comprehensive interpretation and quantitative model of the line width variation that is based on our recent {\em Hinode} results and one that intrinsically connects very dynamic spicules with \alfven{} waves on very small spatial scales.

\begin{figure}
\plotone{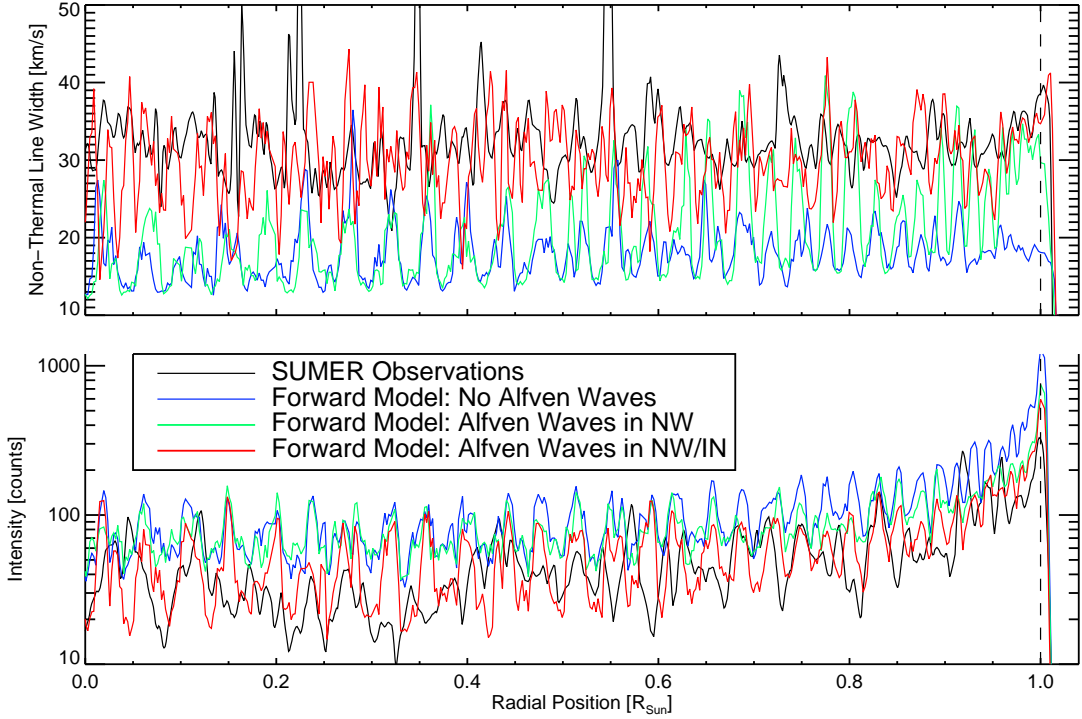}
\caption{Forward modeling of the \ion{C}{4} 1548\AA{} $v_{1/e}$ (top) and intensity (bottom) for the 2\arcsec{} wide synthetic strip of Sun compared to a SUMER observation from a single pixel strip of the spectroheliogram shown in Fig.~\pref{f1} (black lines). We show the results of three identical feature distributions: no \alfvenic{} motions at all (blue), \alfvenic{} motions only in the network features (green) and \alfvenic{} motions on all features (red). This model has the following parameters: $\theta_{I,II}=0\arcdeg$, $\theta_{C}=90\arcdeg$, $\sigma_\theta=30\arcdeg$, ${h_l}_I=300$km, ${h_l}_{II}=4000$km, ${h_l}_{IN}=3000$km, ${h_s}_{I,II}=1500$km, ${h_s}_{C}=2000$km, ${v_l}_{I,C}=0$km/s, ${v_l}_{II}=40$km/s, ${\sigma_l}_I=20$km/s, ${\sigma_l}_{II}=40$km/s, ${\sigma_l}_{C}=10$km/s, $v_t=60$km/s, $\sigma_t=5$km/s, ${h_i}_{I,II}$=2000km, ${h_i}_{C}$=3000 km. We distribute $n_I=571$ type I spicules and $n_{II}=1540$ type II spicules over 43 network regions, with $n_{C}=6899$ structures in between the network regions. The network regions are set to occur periodically every
30 Mm and can be recognized by the sharp peaks in intensity and
linewidth of the blue line in Fig.~\ref{f4}. The internetwork regions
are the regions in between the network locations.\label{f4}}
\end{figure}

We have developed a forward model to test whether the observed on-disk and limb behavior of the \ion{C}{4} line can be explained by LOS superposition of unresolved \alfvenic{} and longitudinal motions of type I and II spicules. We adopt a Monte Carlo approach to simulate spectral profiles based on the superposed motion of spicules and of internetwork, nearly-horizontal, ``canopy''-like ($C$) structures \--using the SOT observations of the statistical properties of these features as a strong observational constraint. We allow a large number of structures ($n=9000$) to occupy a narrow (2\arcsec{} wide) strip of the Sun's surface from disk center to 20\arcsec{} above the limb, prescribing that $n_I$ type I and $n_{II}$ type II spicules occur in close proximity to network elements (given a constant separation of $30$Mm) and $n_{C}$ independent $C$ structures that connect those network elements. The SOT observations \citep{DePontieu2007c,DePontieu2007b} give constraints for each feature (spicule or $C$ structure): an angle with respect to the vertical (randomly chosen from a Gaussian around $\theta$ with 1/e width $\sigma_\theta$; $G(\theta;\sigma_\theta)$), a length $l$ (randomly chosen from an exponential distribution with scale height $h_l$; E(l; $h_l$)), a fixed starting height above the surface ($h_s$), a longitudinal velocity G($v_l$;$\sigma_l$), a transverse or \alfvenic{} amplitude G($v_t$; $\sigma_t$), the observed \alfven{} wave periods (uniformly chosen between 100 and 500s), and a fixed relative intensity $I$ that varies along its length with scale height $h_{i}$. After selecting the location, angle, intensity and velocities of each feature, we compute the spectrum of the simulated strip under the assumption that the simulated spectral line is optically thin, i.e, the ``observed'' line profile in each pixel is the sum of thermal width Gaussians emitted by each feature (in that pixel) that is then Doppler shifted by the net velocity of the feature projected onto the LOS (i.e., a mixture of the longitudinal and transverse velocities). We take into account the smearing effect of different \alfven{} wave phases during longer exposure times by calculating several spectra per ``exposure''.  Finally, for each pixel in the strip, we perform GA Gaussian fits to deduce the intensity, line width (see Fig.~\ref{f4}) and Doppler shift. We note that the network-related structures (type I and type II) are forced to occur within 5 Mm of the network loci and that the $C$ structures are allowed to occur everywhere in-between. In addition, we note that while spicular motions play an important role in this model, our model does not necessarily imply that all TR emission comes from spicules. After all, the scale height of the type I spicules is set to be rather small (400 km), and a large number of $C$ structures occur in the model. We point out that numerical simulations and observations imply that spicular motions have a significant impact on the observed velocities and line widths of the TR \citep{Hansteen2006,DePontieu2007c}. Whether the spicules (e.g., type I) emit significantly in the UV or not requires further (detailed) investigation, but the same simulations show that the TR emission will be moved up and down by the underlying chromospheric features.

The properties of each feature (type I, II, or $C$) are set by thirteen parameters, most of which are strictly constrained by SOT observations: e.g., scale height, minimum height, velocities, angle from the vertical, period range of the waves, etc. The number of structures per network element or the relative intensities of the network and canopy features are less well constrained. However, we stress that the point of this effort is not to derive such detailed parameters, but rather to investigate qualitatively which features have a significant effect on the network contrast and CL variation of the SUMER line widths. We will discuss the intricacies of the forward model in a future (longer) paper.

Fig.~\ref{f4} shows a comparison of the simulated CL line width (top) and intensity (bottom) variations, compared to a sample of one radial (non-averaged) cut from the SUMER \ion{C}{4} spectra of Fig.~\pref{f1} (black lines). We see that a model in which none of the structures carry Alfv\'en waves (blue lines) shows a significant, but too small, increase in line widths around network elements (from longitudinal spicular flows), and lacks the necessary line width in the internetwork.  More importantly, this model does not show any increase of line width toward the limb, both of which {\em are} observed in SUMER spectra (Fig.~\ref{f2}). By switching on \alfven{} waves with 40km/s rms amplitude \citep[green line; reasonable given the projections of, e.g.,][]{Cranmer2005} in the network spicules, we see that the network line widths are now similar in amplitude to those observed by SUMER. In addition, the superposition of many structures in various network elements along the LOS towards the limb naturally leads to an increase of the line width above 0.9R$_{\sun}$. However, the internetwork line widths are still too low. A final sample model run includes \alfven{} waves in all structures (again with 40km/s rms amplitude; red line) and shows a striking similarity of intensity and line width profiles with those observed by SUMER. It reproduces not only the variability of the intensity across the disk and the average values of the line width, but also reproduces the values in the internetwork region, the generally isotropic nature of the CL line width variation, network enhancement and even the bump that starts just inside the limb and peaks a few arcseconds above it.


\section{Discussion}

Our results indicate that the SUMER \ion{C}{4} observations of line widths are indeed fully compatible with the LOS superposition of a large number of finely scaled spicules that carry significant longitudinal mass flows as well as vigorous \alfven{} waves. Since the longitudinal and transverse velocity components are of roughly the same order, the overall center-to-limb variation is relatively limited, naturally producing an isotropy of line widths which had been used as an argument against \alfven{} waves in the past \citep[e.g.][]{Chae1998}. There is a slight but significant increase of line widths in a region close to and off the limb, which seems to be caused by \alfvenic{} motions along the much increased number of structures (the first few 2\arcsec{} pixels around the limb contain spicules from three neighboring network elements!). The steep drop-off of the line width as we go above a certain height above the limb is directly related to the dwindling number of spicular structures: spicules have typical heights of 5,000 km with very few above 10,000 km. This straightforward interpretation contradicts earlier work that explains this decrease as a result of ion-cyclotron damping of high-frequency \alfven{} waves \citep{Peter2003}. Our model also suggests that the increase of line widths in and around the magnetic network is directly related to the concentration of spicule flows along the field {\em and} \alfvenic{} motions.

We expect that similar results will hold for UV emission lines formed in the low TR (T$<$250,000K) although we note that opacity plays a significant role in the broadening of cooler lines \citep[e.g.,][]{Doschek2004}. It is possible that \ion{C}{4} has some opacity at the limb as well, explaining why the forward model predicts intensity increases at the limb that are twice those observed. 
It is interesting that, for emission lines formed at temperatures above 500,000K (e.g., like those of \ion{Ne}{8}), the bump at the limb does not appear, instead we observe a linear increase of line width with the decreasing density above the limb, indicative of the undamped growth of \alfven{} waves in the extended corona \citep[e.g.,][]{Banerjee1998}. Our model does not apply to these hotter lines, since the spatial distribution of emission from these lines is no longer dominated by spicules, but rather by thermal conduction and the overlying coronal structure.

Further study will be necessary to explain the larger (by $\sim$5km/s) line widths in polar and equatorial coronal holes. These enhancements may well be caused by the lower densities and the resulting higher amplitude of the \alfven{} waves (for an equal energy flux). Detailed studies with SOT are necessary to determine whether different amplitudes for the spicular flows perhaps also play a role, especially since the mix of spicule types changes considerably in coronal holes \citep{DePontieu2007b}. We note that the absolute magnitude of the line width is of significant importance in constraining the amplitude of the \alfven{} waves that are needed for solar wind models. Unfortunately, the literature reveals a wide scatter of values for the same TR UV emission lines \cite [see, e.g.,][]{Mariska1992,Chae1998} from different instruments, epochs, locations, etc. This is perhaps not surprising since our model shows that we can expect significantly different values of line widths for the same line depending on proximity to the limb, the small-scale distribution of magnetic flux (which varies considerably over the solar cycle) and the orientation of the magnetic field (since inclined spicules will show varying contributions of longitudinal and transverse velocities). In addition to these physical causes, we have found that the fitting algorithm and form of functional fit (Gaussian with constant, linear or quadratic background) can have a significant impact ($\sim 10$ km/s) on the magnitude of the line width; e.g., the more complex the background form of the fit the smaller the resulting line width.

Our forward model is only a first, primitive, step towards a comprehensive understanding of what dominates the emission of the TR. There has been extensive discusson on the detailed physical nature of the TR in the past \citep[e.g.,][]{Judge1999,Peter2000a}, much of which has ignored the intrinsically dynamic nature of the TR \citep[see, e.g.,][]{Wikstol1998}, which is inevitable given its connection to spicules. It is clear from our results that more sophisticated models of the TR are needed that take into account the intrinsically dynamic and finely structure nature of its dominant components.

\acknowledgements 
SWM was supported by NSF grant ATM-0541567, NASA grant NNG06GC89G and internal funding from SwRI (R9720). BDP by NASA grants NAS5-38099 ({\em TRACE}), NNM07AA01C ({\em Hinode}) and NNG06GG79G. {\em Hinode} is a Japanese mission developed and launched by ISAS/JAXA, with NAOJ as a domestic partner, NASA and STFC as international partners.

\clearpage


\end{document}